# REVIEW ON FEATURE SELECTION TECHNIQUES AND THE IMPACT OF SVM FOR CANCER CLASSIFICATION USING GENE EXPRESSION PROFILE

[1]G.Victo Sudha George  and  [2]Dr. V.Cyril Raj

[1] Asst.Professor ,Dept of CSE ,  Dr.M.G.R Universiy,Chennai-95 ,India
`sudhajose72@yahoo.com`
[2] Professor& Head ,Dr. M.G.R University ,Chennai -95,India

*ABSTRACT*

*The DNA microarray technology has modernized the approach of biology research in such a way that scientists can now measure the expression levels of thousands of genes simultaneously in a single experiment. Gene expression profiles, which represent the state of a cell at a molecular level, have great potential as a medical diagnosis tool. But compared to the number of genes involved, available training data sets generally have a fairly small sample size for classification. These training data limitations constitute a challenge to certain classification methodologies. Feature selection techniques can be  used to extract the marker genes which influence the classification accuracy effectively by eliminating the un wanted noisy and redundant genes This paper presents a review of feature selection techniques that have been employed in micro array data based cancer classification  and  also the predominant role of SVM for cancer classification.*

## 1. INTRODUCTION

DNA micro array is a prominent high throughput technology that allows the expression levels of thousands of genes to be monitored simultaneously. Today, the analysis of gene expression data is one of the major topics in health informatics [1]. For instance, the classification of DNA micro array data allows the discovery of hidden patterns in expression profiles and opened possibility for accurate cancer classification.

The main challenge in classifying gene expression data is the curse of dimensionality problem. There is large number of genes (features) compared to small sample sizes [2,3]. To overcome this, feature selection is used to identify differentially expressed genes and to remove irrelevant genes. Gene selection remains as an critical task to improve the accuracy and speed of classification systems[4].In general, feature selection can be organized into three categories: filter, wrapper and embedded methods. They are categorized based on how a feature selection technique combines with the construction of a classification model. A considerable amount of literature has been published on gene selection methods for building effective classification model. In this paper we present a review of feature selection techniques for cancer classification and also the predominant role of SVM for cancer classification.

## 2. DNA MICROARRAY

Microarrays offer an efficient method of gathering data that can be used to determine the expression pattern of thousands of genes. The mRNA expression pattern from different tissues in normal and diseases states could reveal which genes and environmental conditions can lead to disease. The experimental steps of typical microarray began with extraction of mRNA from a tissues sample or probe. The mRNA is then labeled with fluorescent nucleotides, eventually yielding fluorescent (typically red) cDNA. The sample later is incubated with similarly processed cDNA reference (typically green). The labeled probe and reference are then mixed and applied to the surface of DNA microarrays, allowing fluorescent sequences in the probe-reference mix to attach to the cDNA adherent to the glass slide. The attraction of labeled cDNA from the probe and reference for a particular spot on microarray depends on the extent to which the sequences in the mix (probe - reference) complement the DNA affixed to the slide. A perfect compliment, in which a nucleotide sequence on a strand of cDNA exactly matches a DNA sequence affixed to the slide, is known as hybridization. Hybridization is the key element in microarray technology. The populated microarray is then excited by a laser and the consequential fluorescent at each spot in the microarray is measured. If neither the probe nor the reference samples hybridize with the gene spotted on the slide, the spot will appear in the black color. However, if hybridization is predominantly with the probe, the spot will be in red (Cy5). Conversely, if hybridization is primarily between the reference and DNA affixed to the slide, the spot will fluoresce green (Cy3). The spot can also incandescent yellow, when cDNA from probe and reference samples hybridize equally at a given spot, indicating that they share the same number of complementary nucleotides in particular spot. Using image processing software, the red-to-green fluorescence will be digitized and providing the ratio values output indicating the expression of genes. The process of microarray experiment is illustrated in Figure 1.

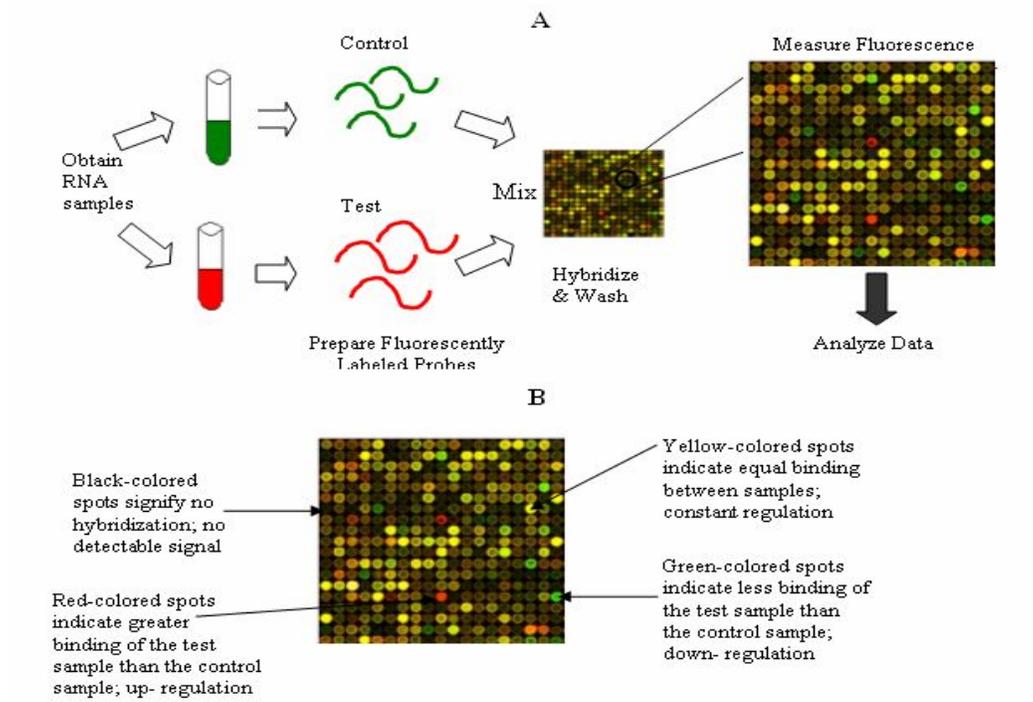

**Figure 1: Microarray Experiment**

Finally, the gene expression data set can be noted by the following matrix $M \{ w_{ij} | 1 \leq i \leq n, 1 \leq j \leq m \}$, where the rows ($G \{ g_1, \ldots, g_n \}$) from the expression patterns of genes, the columns ($S \{ s_1, \ldots, s_m \}$) from the expression profiles of samples, and $w_{ij}$ is the measured expression level of gene $i$ in sample $j$. Thus, M is defined as:

$$M = \begin{bmatrix} w_{11} & w_{12} & \cdots & w_{1m} \\ w_{21} & w_{22} & \cdots & w_{2m} \\ \vdots & \vdots & \ddots & \vdots \\ w_{n1} & w_{n2} & \cdots & w_{nm} \end{bmatrix} \leftarrow g_i, i = 1, \cdots, n$$

$$\uparrow$$
$$s_j, j = 1, \cdots, m.$$

Due to its high throughput nature, microarray data poses new challenges for data analysis. Although the type of analysis depends on the research questions posed, typical steps in the analysis of microarray data are: i) pre-processing and normalization, ii) detection of genes with significant fold changes, iii) classification and clustering of expression profiles.

## 3. CHALLENGES FACED IN MICROARRAY DATA ANALYSIS

Many challenges in microarray need to be addressed before new knowledge about gene expression can be revealed. Some of the problems are:

**a.** Bias and confounding Problem: which occurred during study, design phase of microarray and can lead to erroneous conclusion. Technical factors, such as differences in physical, batch of reagents used and various levels of skill in technician could possibly cause bias. Confounding on the other hand, take place when another factors distorts the true relationship between the study variables of interest.

**b.** Cross-platform comparisons of gene expression studies are difficult to conduct when microarrays were constructed using different standards. Thus, the results cannot be reproduced. To deal with this problem, Minimal Information about a Microarray Experiment (MIAME) [5] has been developed to improve reproducibility, sensitivity and robustness in gene expression analysis.

**c.** Microarray data is high dimensional data characterized by thousand of genes in few sample sizes, which cause significant problems such as irrelevant and noise genes, complexity in constructing classifiers, and multiple missing gene expression values due to improper scanning. Moreover, most of studies that applied microarray data are suffered from data over fitting, which requires additional validation.

**d.** Mislabeled data or questioned tissues result by experts also another types of drawback that could decrease the accuracy of experimental results and led to imprecise conclusion about gene expression patterns.

**e.** Biological relevancy result is another integral criterion that should be taken into account in analyzing microarray data rather than only focusing on accuracy of cancer classification. Although there is no doubt gaining high accuracy classification results are important in microarray data analysis, but revealing the biological information during the process of cancer classification is also essential. For instance determination of genes that are under expressed or over expressed in cancerous cells could assists domain experts in designing and planning more appropriate treatments for cancer patients. Therefore, most of domain experts are interested in classifiers that not only produce high classification accuracy but also reveal important biological information.

## 4. FEATURE SELECTION TECHNIQUES IN MICRO ARRAY DATA ANALYSIS

DNA micro array technology is used to measure changes in expression levels of genes. This expression of the genetic information occurs in two stages: transcription stage and translation stage. In transcription, DNA molecules are transcribed into mRNA while in translation stage, mRNA is translated to amino acid sequences of the corresponding proteins. DNA micro array analysis provides access to thousands of genes at once by recording expression levels simultaneously. It has been shown that gene expression changes are related with different types of cancers [37]. Cancer classification using gene expression data is a nontrivial task due to the very nature of the gene expression data. The expression data has very high dimensionality, usually in the order of thousands to tens of thousands of genes. The situation is more complicated with the number of sample sizes, usually below hundred. The high dimensionality of the features and the low population size usually cause over-fitting of the classifier. A term - the curse of dimensionality, is coined to refer to this situation. Computational expenses also impose important limitations. Another key issue is, due to not all genes being related to the cancer, it is difficult to extract biologically meaningful genes.

The Taxonomy of dimensionality reduction techniques can be divided into two categories, transformation or selection based reduction. The key distinction made within the taxonomy is whether a dimensionality reduction technique will transform or preserves the dataset semantics in the process of reduction. Transformation based reduction such as Principal Component Analysis (PCA) transforms the original features of a dataset with a typically reduced number of uncorrelated ones, termed principal component. In contrast, selection reduction techniques attempt to determine a minimal feature subset from a problem domain while retaining the meaning of the original feature sets. Thus, selection based reduction techniques have become the main preference in many bioinformatics applications, especially microarray data analysis since it offers the advantage of interpretability by a domain expert. Feature selection is the process of systematically reducing the dimensionality of a dataset to an optimal subset of attributes for classification purposes. Problem of feature selection is hence, an important issue in cancer classification. It has been shown that, in many applications feature selection process improves a classifier's prediction capability [38].

The objectives of feature selection techniques are many, the major ones are: i. To avoid over fitting and improve model performance, for example selecting highly informative genes could enhance the accuracy of classification model. ii. To provide faster and more cost-effective models, and iii. To gain a deeper insight into the underlying processes that generated the data. Although, feature selection techniques have many benefits, it also introduces extra complexity level which requires thoughtful experiment design to address the challenging tasks, yet provide fruitful results. In the context of classification, feature selection techniques can be organized into three categories, depending on how they combine the feature selection search with the construction of the classification model: filter method, wrapper method and embedded method.

Filter method rank each feature according to some univariate metric, and only the highest ranking features are used while the remaining low ranking features are eliminated. This method also relies on general characteristics of the training data to select some features without involving any learning algorithm. Therefore, the results of filter model will not affecting any classification algorithm. Moreover, filter methods also provide very easy way to calculate and can simply scale to large-scale microarray datasets since it only have a short running time. Univariate filter methods such as Bayesian Network [6] Information Gain (IG) and Signal-to-Ratio(SNR) [7][10] and Euclidean Distance [8][9] have been extensively used in microarray data to identify informative genes. Information Gain has been reported to be the superior gene selection technique by [8][11] however different types of univariate technique appears to be significant when it was trained over various datasets. Bayesian Networks, on the other hand appear to be the ideal platform for the integration of heterogeneous sources of information [6]. Beside the application of parametric techniques in determining informative genes from microarray data [12][3][14] have applied non-parametric technique such as threshold number of misclassification or TNoM score. This technique basically separate the informative gene by assigning a threshold value. However, it is hard to determine the most appropriate threshold. Other nonparametric techniques such as Pearson correlation coefficient [8] [9] and Significant Analysis of Microarray (SAM) [15] as been reported to be the top feature selection techniques. Univariate filter methods have been widely utilized in microarray data analysis. This trend can be clarified by a number of reasons for instance the output the result provide by univariate gene rankings is intuitive and easy to understand. These simplify version of output could fulfill the aims and expectations of biology and molecular domain experts who demand for validation of result using laboratory techniques. In addition, filter methods also offer less computational time to generate results which is an extra point to be preferred by domain experts. However, gene ranking based on univariate methods has some drawbacks. The major one is the genes selected are most probably redundant. This means highly ranked genes may carry similar discriminative information towards the defined class. Although we eliminate one high ranked gene it may not cause any degradation of classification accuracy. Since univariate filter methods do not count the relationship between genes, [16] developed an optimal gene selection method called Markov Blanket Filtering, which can remove redundant genes to eliminate this problem. Based on this method [17] proposed the Redundancy Based Filter (RBF) method to deal with redundant problems and the results are quite promising.

While the filter techniques handle the identification of genes independently, a wrapper method on the other hand, embeds a gene selection method within a classification algorithm. In the wrapper methods [18] a search is conducted in the space of genes, evaluating the goodness of each found gene subset by the estimation of the accuracy percentage of the specific classifier to be used, training the classifier only with the found genes. It is claimed that the wrapper approach obtains better predictive accuracy estimates than the filter approach [19] however, its computational cost must be taken into account. Wrapper methods can be divided into distinct groups, deterministic and randomized search algorithm. Genetic Algorithm (GA) is a randomized search algorithm and optimization mimicking evolution and natural genetics. It has been employed for binary and multi-class cancer discrimination in [20][21]. A common drawback of wrapper methods, such as GA is that they have a higher risk of over-fitting than filter techniques and are very computationally intensive. In contrast, wrapper methods incorporate the interaction between genes selection and classification model, which make them unique compared to filter techniques.

The third class of feature selection approaches is embedded methods. The different of embedded methods with others feature selection methods is the search mechanism is built into the classifier model. Identical to wrapper methods, embedded methods are therefore specific to a given learning algorithm. Embedded methods have the advantage that they include the interaction with the classification model, while at the same time being far less computationally intensive than wrapper methods. Support Vector Machine (SVM) method of Recursive Feature Elimination (RFE) was employed in [22] for gene selection.

## 5. SUPERVISED CLASSIFICATION AND SVM

Supervised classification, also called prediction or discrimination, involves developing algorithms to priori defined categories. Algorithms are typically developed on a training dataset and then tested on an independent test dataset to evaluate the accuracy of algorithms. Support vector machines are a group of related supervised learning methods used for classification and regression. The simplest type of support vector machines is linear classification which tries to draw a straight line that separates data with two dimensions. Many linear classifiers (also called hyperplanes) are able to separate the data. However, only one achieves maximum separation. Vapnik in 1963 proposed a linear classifier as a original optimal hyper plane algorithm [24]. The replacement of dot product by a nonlinear kernel function allows the algorithm to fit the maximum-margin hyperplane in the transformed feature space [23,24]. SVM finds a linear separating hyperplane with the maximal margin in this higher dimensional space is called the kernel function [24]. There are four basic kernels: linear, polynomial, radial basic function (RBF), and sigmoid [25].

## 6. WHY SVM FOR CANCER CLASSIFICATION

Gene expression Microarrays are becoming increasingly promising for clinical decision support in the form of diagnosis and prediction of clinical outcomes of cancer and other complex diseases. In order to maximize benefits of this technology, researchers are continuously seeking to develop and apply the most accurate decision support algorithms for the creation of gene expression patient profiles. Prior research suggests that among well-established and popular techniques for classification of microarray gene expression data, support vector machine(SVMs) achieve the best classification performance, significantly out performing K-nearest neighbors, back propagation neural networks, probabilistic neural networks, weighted voting methods and decision trees.

The reasons for this are

1. SVMs have demonstrated he ability to not only correctly separate entities into appropriate classes, but also to identify instances whose established classification is not supports by the data.

2. SVM have many mathematical features that make them attractive for gene expression analysis, including their flexibility in choosing a similarity function, sparseness of solution when dealing with large data sets, the ability to handle large feature spaces, and the ability to identify outliers.

## 7. RELATED WORK

In 1999 SVM-based method for directly classifying genes based on microarray data [26] was perhaps first published. In that features were used along with both the polynomial and Gaussian kernel and SVM were trained to distinguish between six functional classes, and their performance were compared with that of four other standard algorithms: Parzen windows, Fisher's linear discriminant, and C4.5 and MOC1 decision trees. The accuracy of both kernels, in particular the Gaussian kernel, surpassed those of all four alternative machine learning techniques in terms of overall error rate. Another early application of SVMs to microarray data was that of tissue classification, one that remains popular today. Numerous studies released around the same time all achieved similarly encouraging results: In 1999 [27] provided the first tissue classification algorithm using SVMs, applied to the problem of differentiating between two types of leukemia. In that feature set was selected based on the signal-to-noise ratio. The same feature set was used by [26] in 2000 for classification of ovarian cancer tissues. Both implementations outperformed Naïve Bayes and other standard machine learning techniques that were typically used for these tasks. The applications of SVMs to microarray data continue to develop and achieve higher accuracy and robustness.

One particularly intriguing innovation is that in 2008 SVMs are used to classify the pixels themselves, either into two groups (foreground and background) or three (signal, background and artifact) [28]. This type of partitioning is typically done using clustering and other unsupervised machine learning algorithms, but the implementation manages to achieve extremely high accuracy. The pixels themselves are represented by vectors of eleven distinct features, which measure the intensity of the pixel, the intensity of its neighbors, and the variation within its neighbors, among others. The classifiers are trained on already-classified data, and are tested on real new microarrays and simulated microarrays, along with various types of preprocessing filters. In every case, the sensitivity of the classifiers exceeds 98%; the accuracy and specificity exceed 99:8%. Recently, another type of classifier has been gaining attraction in microarray classification, namely random forests. Random forests are another type of machine learning algorithm which consist of many randomly-generated (by a bootstrap-like process) decision trees. The output of the classifier for a given input is the most popular result among all the random trees. In July 2008 [29] released a comparison of random forests and SVMs for classifying cancer tissue based on microarray data. According to the authors, random forests, despite their increasing popularity, are still not as accurate as SVMs for typical microarray classification problems. The metric for the comparison was the area under the respective ROC curves, as well as the relative classifier information (RCI), an entropy-based measure that can be applied to multi-class decision problems. By these measures it is proved that SVMs out formed random forests on nine of eleven and eleven of eleven tasks, respectively. SVMs have demonstrated the ability not only correctly separate entities into appropriate classes, but also to identify mis-labeled data [31].

Another interesting and recent development is due to [30] in 2009, approach is the same standard binary classification problem as previous researchers, but incorporate network-based information into the training programs. More specifically, an underlying model of statistically significant subnetworks is constructed by searching various subnetworks and assigning scores based on each subnetwork's gene expression level; the algorithm then identifies subnetworks that are capable of discriminating effectively between categories. Using this information, a penalty term is constructed which penalizes contradictions between some classification and the corresponding subnetwork model(s). This penalty term is added to the optimization program itself, rather than incorporated into the feature space or the kernel function, but the effect is still that the resulting classifiers are biased towards the underlying subnetwork models. According to the authors, this technique improves the consistency of the SVM classifier, and also allows for the extraction of higher-level biological data which is available in other databases and formats. In [32] performance of SVM is investigated with linear regression and neural network on colon tumor data sets after performing feature selection. 10 and 50 features were selected by t-statistic feature selection method and achieved maximum of 85% accuracy on SVM with RBF kernel. In [33] a novel feature selection method namely recursive feature elimination (RFE) introduced and experiments were done on colon tumor and leukemia gene expression dataset. With the colon cancer dataset using 4 genes the method used achieved 98% accuracy.

In [34] Eight data sets used in the experiment and almost in all cases, the accuracy and performance of classifiers were improved after applying feature selections methods to the datasets. In all cases SVM-RFE performed very well when it applied with SVM classification methods. In lymphoma dataset SVM-RFE performed 100% in combination of SVM classification method. In [35] authors used 10 published microarray datasets, encompassing 6 binary and 4 multiclass classification Problems and conducted a comprehensive study of both classification methods as well as feature selection methods for classification of microarray data. All implementations of machine learning algorithms were taken from the Weka library. Therefore assumed an approximately equal quality of implementations and differences can be attributed to the methods themselves and not to implementations. The experiments focused on identifying the best combination of classifier and feature selection strategy. For this purpose, a selection of common (esp. three SVM variants) and less common classifiers (such as Voted Perceptron and One Rule) was trained on a large variety of feature selections, produced by both wrapper and filter strategies. Leave-one-out cross-validation is used for evaluation and altogether around 220,000 different combinations of classifiers and feature selections were analyzed. As a general result, it was found that linear SVM to be the best classifier in the field, closely followed by the quadratic kernel SVM. It is reported that classification using the SVM method can be improved using a well-chosen size of features together with problem-dependent feature selection techniques. For classifiers, the linear and quadratic SVMs show the best overall performance.

It is Demonstrated that SVM can not only classify new samples, but can also help in the identification of those which have been wrongly classified by experts [36]. SVMs are unique among classification methods in this regards.

## 8. CONCLUSION AND FUTURE DIRECTIONS

This paper reviewed first, the feature selection techniques that have been employed in cancer classification using gene expression profiles. High dimensionality input and small sample data size are the main two problems that have been triggers the application of feature selection in microarray data analysis. Numerous and fruitful efforts have been conducted during the past several years in the utilization of feature selection to encounter these problems, which mainly can be grouped into three main approaches; filter, wrapper and embedded approaches. And it is seen that SVM-RFE is now gaining popularity. Secondly the predominant performance of SVM for Cancer classification is reviewed. Support Vector Machine (SVM) has recently gained wide popularity among machine learning community due to its robust mathematical basis and high prediction performance. It has been successfully applied to the wide variety Cancer classification problems .And it is seen that a feature selection method called SVM-RFE performed very well when it applied with SVM classification methods and it is demonstrated that for lymphoma dataset, SVM-RFE performed 100% in combination of SVM classification [34].

A considerable amount of literature has been published on gene selection methods for building effective classification model. However, a large part of these literatures are statistical analysis, and their algorithms consider solely on gene expression values to select optimal feature subset. Although these have shown a promising classification results but there are still some disadvantages on them. The expression values may not be accurately measured and the complexity of micro array experiments can causes discrepancy in data obtained. Moreover statistical significance might not able to directly translate into biological relevance. In recent years, researches have realized that gene markers identified from microarray drawn from different studies on the same disease across similar cohorts lack consistency [39,40]. And in the past few years, study on integrative analysis on micro array data, which is described by [14] as the analysis of high throughput data in the context of available biological knowledge is gaining popularity.

Recently efforts are directed towards integrative gene selection methods that consider gene expression data along with additional biological information like Gene Ontology and metabolic and regulatory pathways (example the MetaCyc and KEGG pathway databases) [41][42][43][44][45].

## ABOUT AUTHORS

**[1] Mrs. G.Victo sudha George** had completed B.E(CSE) in the year 1993 and M.Tech in the year 2007. Currently pursuing Ph.d., and the area of research is Bioinformatics. Other areas of interest are MobileComputing, Data Mining and Artificial Intelligence At present working as Assistant professor in the Departmentof Computer Science and Engineerng in Dr.M.G.R University ,Chennai,India.

**[2] Dr. V.Cyril Raj** had completed M.E, and PhD and his areas of interest are Bioinformatics Data Mining,Grid Computing,Mobile Computing,Robotics etc.He had published a lot of papers in International and National level journalsand also authored many books.. At present working as Professor and Head of Computer science and Engg. Dept in Dr.M.G.R University,Chennai,India.